\documentclass[preprint2]{aastex}

\pdfoutput=1

\usepackage{natbib}
\usepackage{amsmath}
\usepackage{times}
\usepackage{graphicx}
\usepackage{epstopdf}
\usepackage{textcomp}
\usepackage{color}

\slugcomment{Accepted for publication in ApJL 2014}

\shorttitle{Water Vapor in Barnard 5}
\shortauthors{Wirstr\"om et al.}

\begin{document}
\title{COLD WATER VAPOR IN THE BARNARD 5 MOLECULAR CLOUD}

\author{E.~S.~Wirstr\"om$^1$, S.~B.~Charnley$^2$, C.~M.~Persson$^1$, J.~V.~Buckle$^{3,4}$, M.~A.~Cordiner$^{2,5}$, and S. Takakuwa$^6$}
\affil{$^1$Department of Earth and Space Sciences, Chalmers University of Technology, 
Onsala Space Observatory, SE-439 92 Onsala, Sweden; eva.wirstrom@chalmers.se}
\affil{$^2$Astrochemistry Laboratory and The Goddard Center for Astrobiology, Mailstop 691, NASA Goddard Space Flight Center, 8800 Greenbelt Road, Greenbelt, MD 20770, USA} 
\affil{$^3$Astrophysics Group, Cavendish Laboratory, J J Thomson Avenue Cambridge, CB3 0HE} 
\affil{$^4$Kavli Institute for Cosmology, Cambridge, Madingley Road Cambridge CB3 0HA, UK}
\affil{$^5$Institute for Astrophysics and Computational Sciences, The Catholic University of America, Washington, DC 20064, USA}
\affil{$^6$Institute of Astronomy and Astrophysics, Academia Sinica, P.O. Box 23-141, Taipei 106, Taiwan}

\begin{abstract}
After more than 30 years of investigations, the nature of gas-grain interactions at low temperatures remains an unresolved issue in astrochemistry. Water ice is the dominant ice found in cold molecular clouds, however, there is only one region where cold ($\sim$10 K) water vapor has been detected -- L1544. This study aims to shed light on ice desorption mechanisms under cold cloud conditions by expanding the sample.
   The clumpy distribution of methanol in dark clouds testifies to transient desorption processes at work -- likely to also disrupt water ice mantles. Therefore, the \textit{Herschel} HIFI instrument was used to search for cold water in a small sample of prominent methanol emission peaks.
   We report detections of  the ground-state transition of $o$-H$_2$O ($J\!=\!1_{10}\!-\!1_{01}$) at 556.9360 GHz toward two positions in the cold molecular cloud Barnard 5. The relative abundances of methanol and water gas support a desorption mechanism which disrupts the outer ice mantle layers, rather than causing complete mantle removal.
\end{abstract}

\keywords{astrochemistry --- ISM: individual objects (Barnard 5) --- ISM: molecules --- stars: formation --- astrochemistry --- submillimeter: ISM}


\section{INTRODUCTION}

Recent observations of interstellar H$_2$O by \textit{Herschel}\footnote{\textit{Herschel} is an ESA space observatory with science instruments provided by the European-led Principal Investigator consortia and with important participation from NASA.}  have revolutionized our understanding of the physics and chemistry of water at the elevated temperatures ($\sim$100--3000~K) encountered in star-forming gas \citep[e.g.][]{KristensenDishoeck11,Dishoecketal11}.  In these energetic environments the observed H$_2$O molecules could be formed in endoergic reactions involving H$_2$ or readily evaporated and/or sputtered from dust mantles. 

As water is the dominant ice component in dark molecular clouds \citep[e.g.,][]{Oberg11}, measuring its gas-phase abundance can shed light on the nature of the gas-grain interaction at low temperatures ($\sim$10~K) -- an issue that remains unresolved after more than 30 years of investigation \citep[e.g.,][]{Leger85}.  Water vapor has only been detected in one dark cloud core, -- L1544 \citep{Caselli12}, and its low measured abundance supports the low upper limits previously obtained by SWAS \citep{Snell00} and ODIN \citep{Klotz08}. This region is ostensibly the least viable target in which to search for water as many gas-phase molecules are frozen out onto the dust. \citet{Caselli12} attribute the presence of H$_2$O  to the photodesorption of ice molecules by a weak, ambient ultraviolet (UV) radiation field \citep[e.g.,][]{PrasadTarafdar83}. 

However, this UV photodesorption mechanism cannot explain the fact that emission from putative photodesorbed molecules in dark clouds is known to be clumpy and not correlated to density structures as one would expect.  Methanol in particular is a molecule that can only form by grain-surface reactions, and molecular maps show that CH$_3$OH is enriched in some clumps but not in others \citep[e.g.,][]{Buckle06}, suggesting that ice mantles must have been removed by a transient process. Because water and methanol have similar physisorption  binding energies \citep{SandfordAllamandola93}, any such  process  should  therefore also release water molecules.
 The gas phase dissociative recombination of protonated water re-forms water and OH molecules (which can go on to form water again), while dissociative recombination predominantly breaks the C--O bond of protonated methanol, and so the post-desorption lifetime of H$_2$O molecules is much longer than that of CH$_3$OH ($>$$10^4$~years longer at densities of 5$\times$10$^4$~cm$^{-3}$).  
  
In this Letter, we report on the detection of water at two positions in the cold dark cloud Barnard 5 with \textit{Herschel} and discuss the implications for the gas-grain chemistry of molecular clouds.

\section{OBSERVATIONS AND DATA REDUCTION}

\subsection{Barnard 5}
The Barnard 5 cloud in Perseus (B5) has four known protostars, of which the most prominent Class I protostar IRS1 drives an energetic outflow into the surrounding cloud material \citep{Langer96}. CO and NH$_3$ maps indicate that B5 has a very clumpy morphology \citep{Goldsmith86,Pineda11} and molecular maps in several other species (e.g., N$_2$H$^+$, CH$_3$OH, HC$_3$N, SO, CCS, c-C$_3$H$_2$) show that these clumps are chemically differentiated (S. B. Charnley et al. 2014, in preparation) in a manner similar to that found in the sample of \citet{Buckle06}. 
We selected two positions for the $o$-H$_2$O search: a main methanol peak - hereafter the "methanol hotspot" -- that is offset from IRS1 by ($-2$\arcmin,$+5$\arcmin) and not associated with a core or column density peak as traced by C$^{18}$O or 850~$\mu$m continuum \citep{Hatchell05b}; and a weaker secondary peak closer to IRS1, see Figure~\ref{Fig_sixspec}. 

The HIFI instrument \citep{deGraauw10} on-board the \textit{Herschel Space Observatory} \citep{Pilbratt10} was used to observe the ground-state transition of $o$-H$_2$O ($J=1_{10}-1_{01}$) at 556.9360 GHz toward two positions in B5 -- the methanol hotspot at R.A.=3$^{\rm h}$47$^{\rm m}$32\fs10, decl.=+32\degr56\arcmin43\farcs0 and the second methanol peak at 3$^{\rm h}$47$^{\rm m}$40\fs4, +32\degr52\arcmin28\farcs0 (J2000.0) -- on August 11, 2012 (OT2). Position switching mode was used with OFF positions at ($-10$\arcmin,0\arcmin), chosen for lack of CO emission \citep{Goldsmith86}. The LO frequency in HIFI band 1b was set to 564.56 GHz in both L and R polarization, placing the line in the lower side band, and the total integration time (ON+OFF) was about 40 minutes for each source. Both the Wide Band Spectrometer (WBS) and the High Resolution Spectrometer (HRS) were used, and because emission lines from the region are known to be narrow (0.5--0.8 km\,s$^{-1}$) the HRS was configured to high resolution mode with a band width of 230 MHz, corresponding to a spectral resolution of 125 kHz, or $\sim$0.07~km\,s$^{-1}$ at this frequency. The methanol hotspot was also observed for a total of 60 minutes integration time on March 7, 2013 (DDT) using the same instrument setup. Observations were performed under observing programs OT2\_ewirst01\_2 and DDT\_ewirst01\_3, and the data presented here is available from the \textit{Herschel} Science Archive\footnote{http://archives.esac.esa.int/hda/ui/} under observing IDs 1342249424, 1342249425, and 1342266407. 

The beam FWHM is 38$\arcsec$ at this frequency, the forward and main-beam efficiencies $\eta_{\rm l}$\,=\,96\% and $\eta_{\rm mb}$\,=\,75\%, respectively.  Detailed information about the HIFI calibration including beam efficiency, mixer sideband ratio, pointing, etc., can be found on the \textit{Herschel} internet site\footnote{http://herschel.esac.esa.int/}. The in-flight performance is described by \citet{Roelfsema12}.

Spectra of both polarizations were reduced separately using the Herschel Interactive Processing Environment \citep{Ott10}, version 10.0.0. Subsequently, data FITS files were exported to the spectral analysis software XS\footnote{Developed by Per Bergman, Onsala Space Observatory, Sweden; http://www.chalmers.se/rss/oso-en/observations/data-reduction-software} for further reduction and analysis. After linear baseline subtraction and frequency alignment, the two polarizations for each observing ID and spectrometer were averaged together, weighted by system temperature and integration time. Pointing offsets between polarizations were within 7$\arcsec$, i.e., less than 20\% of the beam size.

\subsection{Other Sources}
During the original OT2 program and the subsequent DDT\_ewirst01\_4, we also observed methanol peak positions in L1512, TMC-1 and TMC-1C, all offset from known protostellar activity. 
Table~\ref{Nondettab} summarizes these observations, with WBS rms noise temperatures on the $T_{\rm mb}$ scale. 
Only upper limits on the abundance/emission from  $o$-H$_2$O ($J=1_{10}-1_{01}$)  were obtained. These limits and associated implications are discussed respectively in Sections~\ref{otherAnaSect} and \ref{DiscSect}.

\section{RESULTS AND ANALYSIS}
Two emission peaks, offset from the systemic velocity ($v_{\mathrm{LSR}}$=9.6~km\,s$^{-1}$) at $\sim$8.9~km\,s$^{-1}$ and $\sim$10.3~km\,s$^{-1}$, are detected  toward the methanol hotspot in the combined WBS spectrum from both observing runs. The rms noise at this resolution is 2.9~mK and the confidence level of the detection across a 3~km\,s$^{-1}$ range around the $v_{\mathrm{LSR}}$ is  
6$\sigma$. Comparing the water spectra to those of other molecules toward the methanol hotspot in Figure~\ref{Fig_sixspec}, the water line profile can be interpreted as a self-absorbed emission line around the systemic velocity. The asymmetry in peak intensities, with a suppressed blue component, is then indicative of low velocity expansion or outflow in the methanol clump. This is commonly observed for water in massive star-forming regions \citep[e.g.][]{Ashby00}, but has never before been observed toward cold, non-star forming gas.

Toward the second methanol peak, the WBS spectrum noise rms is 5~mK. We find a tentative detection of water emission at velocities higher than the $v_{\mathrm{LSR}}$, corresponding to 3$\sigma$ rms when integrated over a 3~km\,s$^{-1}$ range around the $v_{\mathrm{LSR}}$, see Figure~\ref{Fig_sixspec}. See below for a further discussion on this tentative detection.

Before further analysis water spectra were multiplied by the total efficiency factor of $\eta_{\rm l}$/$\eta_{\mathrm{mb}}$=1.26 (HIFI Observers' Manual version 2.4) in order to get them on the $T_{\mathrm{mb}}$ scale. Note that the HRS OT2 data show artifacts in the spectrum at the velocity of the emission feature, and is therefore not included in any analysis. 


\subsection{Analysis} \label{ALIsect}
In recent years, submillimeter continuum surveys of the Perseus region have contributed to the understanding of cold dust properties around the B5 methanol hotspot.
A dust temperature map based on \textit{Spitzer} and Two Micron All Sky Survey2 data is available from the COMPLETE database\footnote{http://www.cfa.harvard.edu/COMPLETE/} and estimates $T_{\rm dust}$=13 and 14~K for the methanol hotspot and the second position in B5, respectively \citep{Schnee08}. In addition, the whole Perseus region has been observed with both PACS and SPIRE (S. Pezzuto et al. in preparation) as part of the Gould Belt Survey \textit{Herschel} Key program \citep[KPGT\_pandre\_1,][]{Andre10}. Temperature and column density maps show no column density peak coinciding with the methanol hotspot within the \textit{Herschel} main beam, while the dust temperature agrees with the \citet{Schnee08} map (S. Pezzuto, private communication). The second methanol peak however coincides to within 20\arcsec\ with a column density peak and temperature minimum (12~K), indicating the presence of a core.

The $1_{10}-1_{01}$ emission line becomes optically thick at low column densities, and in cold gas essentially all $ortho$-water molecules reside in their ground-state. However, water excitation and radiative transfer are significantly complicated by the very large optical depths encountered, subthermal excitation, and possibility of population inversion. In order to interpret the water line profile suggested by our observations we use an accelerated lambda iteration (ALI) scheme \citep{RybickiHummer91}, including these non-LTE effects, to solve the radiative transfer in a spherically symmetric model cloud. The ALI code is based on that used and tested by, e.g., \citet{Maercker08}, but modified by P. Bergman to take into account several collision partners. 

To estimate the physical properties of the desorbed gas at the B5 methanol hotspot, the observed methanol emission line quartet around 96.7~GHz (S. B. Charnley et al. in preparation) was modeled using ALI. A- and E-type methanol were modeled separately, with molecular data from the Cologne Database for Molecular Spectroscopy\footnote{\texttt{http://www.ph1.uni-koeln.de/vorhersagen/}} \citep{Muller05} and collision rates adapted from \citet{RabliFlower10}.  Both line shapes and intensities are well reproduced by a homogeneous, 0.08 pc diameter spherical model clump whose physical properties are given in Table~\ref{ALItab}, and where the molecular spin ratio A/E$\sim$1. The size of the methanol emitting region in this model is defined by map data, the total
 H$_2$ column density is of the same order as that found by the Gould Belt Survey, but the gas kinetic temperature has to be lower than their derived dust temperature, as low as 9~K, to reproduce observed line intensity ratios. 
The resulting average column density over the 64\arcsec ARO 12m beam is $N$(CH$_3$OH)=$1.3 \times 10^{14}$~cm$^{-2}$, and the total H$_2$ mass of the model clump is 0.64~$M_{\sun}$, which is less than half of its virial mass (3.02~$M_{\sun}$) indicating that it is gravitationally unbound.

Molecular data for $o$-H$_2$O was taken from the
Leiden Atomic and Molecular Database\footnote{Available at \texttt{http://www.strw.leidenuniv.nl/~moldata}} (LAMDA) \citep{Schoier05}. Collisional excitation of $o$-H$_2$O by both ortho- and para-H$_2$ is taken into account, state-to-state collisional rates are adapted from \citet{Dubernet09} and \citet{Daniel11}, respectively. At low temperatures water excitation is very sensitive to the ortho-to-para ratio (OPR) in H$_2$ and the thermal equilibrium OPR at 10~K is $3.6 \times 10^{-7}$. However, due to formation on grains and chemical recycling of H$_2$, such low OPR is not expected to be reached within the lifetime of a dark cloud \citep[e.g.,][]{Flower06,Pagani11}, which is also consistent with observations \citep{Troscompt09}. Here we adopt a constant OPR value of 0.001, close to that predicted for a 10~K cloud of H$_2$ density 10$^4$~cm$^{-3}$ by \citet{Flower06}.  

\begin{figure*}
   \centering
   \includegraphics[width=8.5cm]{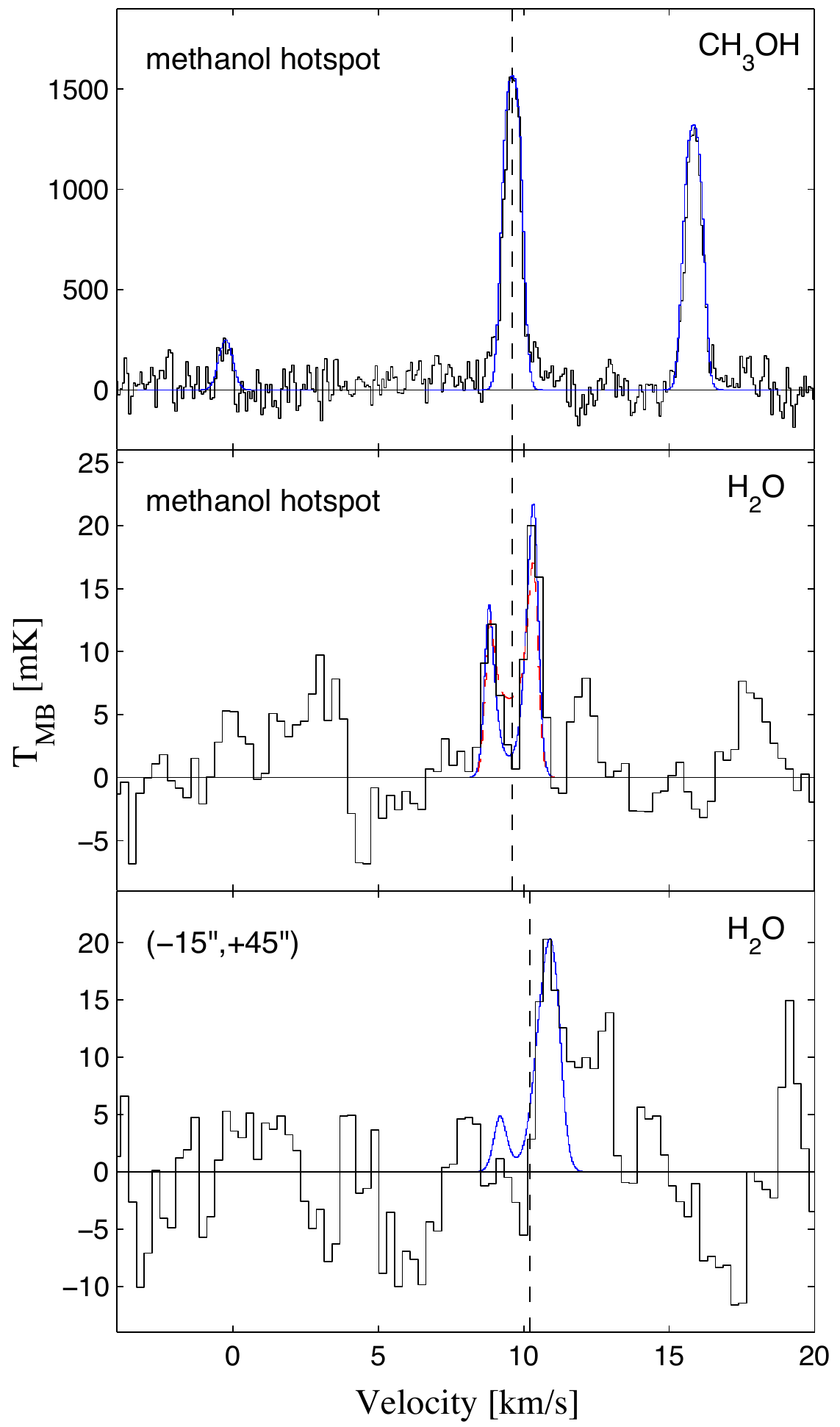}
      \caption{ALI model results of slowly expanding spherical clouds (in blue; same parameters as in Table~\ref{ALItab}) as compared to observed spectral features (black) toward the two positions in B5. Toward the methanol hotspot, the red dashed curve shows modeled water emission when excluding the envelope component.
Local $v_{\mathrm{LSR}}$'s are marked by dashed vertical lines.
 }
         \label{Fig_ALI}
   \end{figure*}

Peak intensities in the water line profile toward the B5 methanol hotspot can be modeled in ALI from the same spherical clump model as used for methanol above by adjusting the turbulent velocity and adding a slow expansion increasing linearly from zero at the cloud center to $v_{\rm exp,max}$ at the cloud edge. This slow expansion velocity can be included in the methanol model without hampering the fit, see the top panel of Figure~\ref{Fig_ALI}. However, in order to reproduce the deep central absorption, an outer envelope of lower density has to be included in the \textbf{water} model. The mid panel of Figure~\ref{Fig_ALI} compares the best fit model with envelope to the corresponding model without, where the temperature, turbulent velocity, and water abundance are kept constant, and the expansion velocity levels out to a constant value in the envelope (parameters in the left column of Table~\ref{ALItab}). Note that in the best-fit model the bulk of the water column originates in the methanol clump while the methanol-free envelope, possibly testifying to a longer time since desorption, only contributes by $\sim$10\%.
At a temperature of 9~K we find that the ortho water abundance, $X_{o-\rm{H}_2\rm{O}}$, has to be around $2 \times 10^{-8}$. At a 10 times higher H$_2$ OPR, this water abundance would only have to be $\sim$5\% lower for the model to fit the data.

Through only small adjustments to the cloud parameters, the water spectrum toward the second methanol peak may also be reproduced using the ALI model. The lower panel of Figure~\ref{Fig_ALI} shows this WBS spectrum together with the model result for a somewhat denser ($n$(H$_2$)=$10^5$~cm$^{-3}$) cloud at $v_{\rm LSR}$=10.2~km\,s$^{-1}$ where the expansion velocity increases to be equal to the turbulent velocity in the outer envelope (see parameters in Table~\ref{ALItab}). This velocity relation is required to explain the lack of a blue-shifted emission component as seen toward the methanol hotspot.  The temperature is kept at 10~K and the ortho water abundance in this model is $5.5 \times 10^{-9}$ relative to H$_2$.

\subsection{Other Sources} \label{otherAnaSect}
For a first-order estimate of the L1512, TMC-1, and TMC-1C water abundance upper limits we exploit the fact that emission in the $o$-H$_2$O ground-state transition never gets thermalized at typical dark cloud densities; therefore, even if the emission is optically thick the high escape probability for emitted photons ensures that we are in the optically thin limit \citep[e.g.,][]{Snell00}. Then the $o$-H$_2$O abundance can be expressed as
\begin{equation}
X(o-{\rm H}_2{\rm O}) = 6.0 \times 10^8 \frac{\int T_{\rm R}~dv}{C(T_{\rm K})\, n_{\rm{H}_2}\,N(\rm{H}_2)}\,\exp (26.8/T_{\rm K}),
\end{equation}
where the integrated intensity is given in K\,km\,s$^{-1}$, the column density in cm$^{-2}$, and the density in cm$^{-3}$. 
Table~\ref{Nondettab} lists the resulting water abundance upper limits, assuming a kinetic temperature of $T_{\rm K}$=10~K, a molecular hydrogen density of 5$\times 10^4$~cm$^{-3}$, and H$_2$ column density of $10^{22}$~cm$^{-2}$.

\section{DISCUSSION} \label{DiscSect}
We have detected cold gas-phase water at two positions in the B5 molecular cloud. The fact that these positions correspond to peaks in the CH$_3$OH distribution supports the view that the water has recently been desorbed from dust grain mantles. The derived H$_2$O/H$_2$ abundance of $\approx 2.5 \times 10^{-8}$  at the methanol "hotspot"  is more than an order of magnitude greater than that found by \citet{Caselli12} in L1544, whereas that derived closer to IRS1 of $\approx 5 \times 10^{-9}$ is comparable. The large amount of water and methanol observed at a position uncorrelated to any core-like density structure suggests that UV photodesorption is not the mechanism for removing molecules from the surfaces of dust grains, but supports the scenario modeled by \citet{CharnleyRodgers09}, where spatial inhomogeneity is a consequence of the ice formation process.

The solid CH$_3$OH fraction toward B5 IRS1 has been estimated to be $<$6\% relative to water ice and so, if evaporated, the water in B5 would have H$_2$O/H$_2$ $\approx$ 10$^{-4}$ \citep{Boogert04}. Even considering a several orders of magnitude lower water ice abundance further away from the cloud density peak, the low observed abundance of gas-phase water at the methanol hotspot implies that the operating desorption mechanism cannot cause complete disruption of ice mantles. 
For a derived CH$_3$OH/H$_2$ abundance of $\approx 4 \times 10^{-8}$  at the methanol hotspot, the gas-phase  CH$_3$OH/H$_2$O ratio in the clump is 1.5.  This demonstrates that only partial removal of ice material rich in methanol has occurred. This is consistent with recent theoretical models of grain mantle formation which follow the chemical composition of each ice monolayer as it is formed \citep{CharnleyRodgers09,GarrodPauly11, VasyuninHerbst13a}. In these calculations CH$_3$OH-rich monolayers form late in the evolution, as the atomic O/H ratio in the gas falls and water formation becomes concomitantly less efficient,  and resides nearest to the mantle surface. Desorption of these monolayers can account for the CH$_3$OH/H$_2$O ratios measured in B5 and therefore places constraints on the efficiency of the actual mantle desorption process. 

\begin{figure*}
   \centering
    \includegraphics[width=\textwidth]{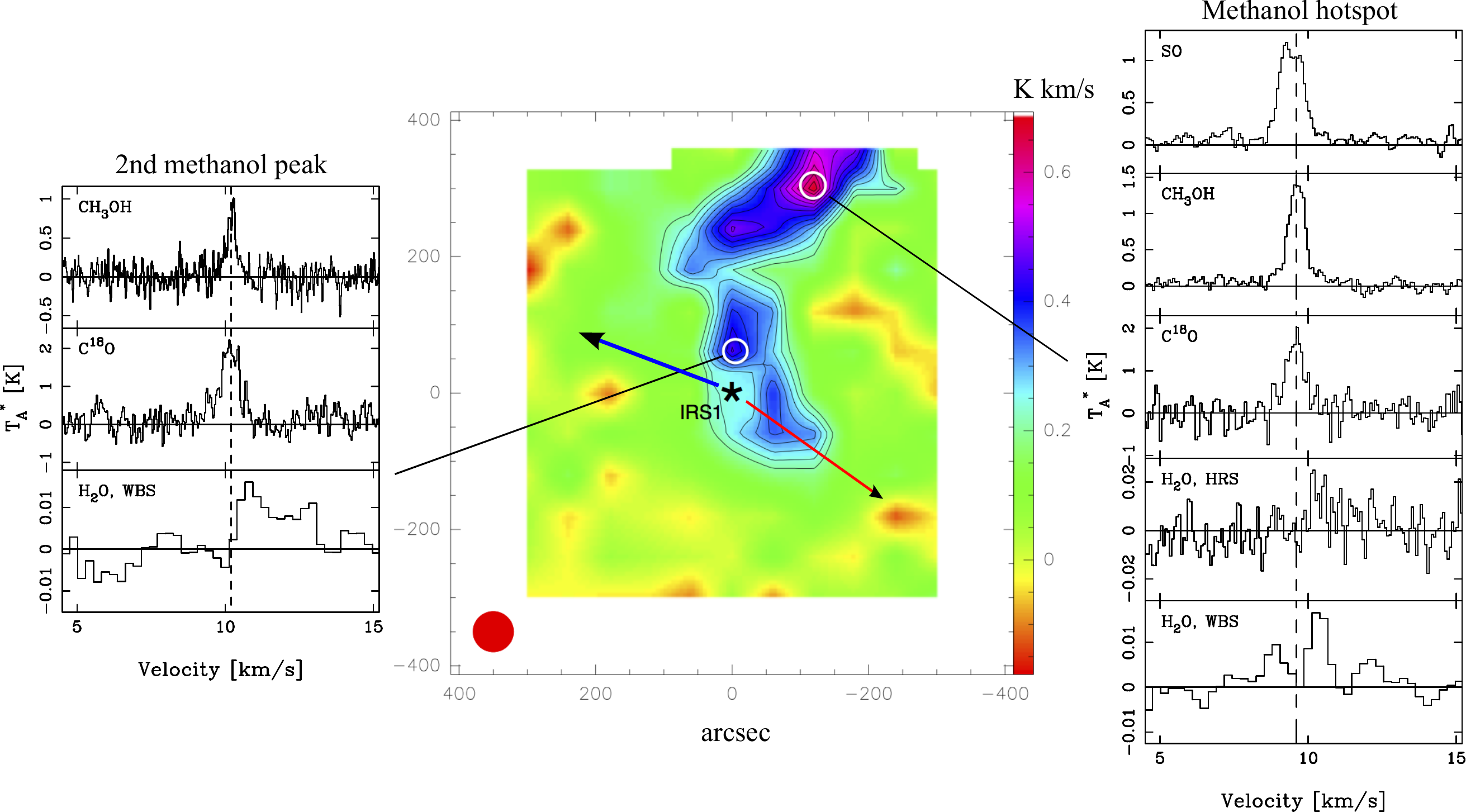}
      \caption{ARO 12 m integrated intensity map of B5 in A-CH$_3$OH at 96.7~GHz together with spectra toward the two observed positions, marked by white rings showing the relative size of the \textit{Herschel} beam. Red and blue arrows show the approximate direction of the IRS1 outflow, spatial scale is in arcseconds from IRS 1. SO, $^{13}$CO and C$^{18}$O were observed with the Onsala 20 m telescope (S. B. Charnley et al. in preparation). Data from both spectrometers (HRS and WBS) are shown, the former redressed to a channel spacing of 0.0756~km\,s$^{-1}$ to correspond to the CH$_3$OH spectral resolution. The vertical dashed lines mark $v_{\mathrm{LSR}}$'s as defined by the CH$_3$OH emission peaks (9.6 and 10.2~km\,s$^{-1}$, respectively).
              }
         \label{Fig_sixspec}
   \end{figure*}

Ice mantle desorption at the methanol hotspot is not likely related to the protostellar activity of IRS 1: the large-scale outflow extends in the NE-SW directions rather than to the NW \citep{Yu99}, and the projected distance from IRS 1 of $\sim$0.55~pc (see Figure~\ref{Fig_sixspec}) is too far for magnetohydrodynamic (MHD) waves to propagate and cause mantle disruption \citep[e.g.,][]{Markwick00}. 
If cosmic-ray-induced photoevaporation or exoergic surface reactions were dominating the desorption, the methanol distribution would be less clumpy and rather follow the H$_2$ column density distribution. Thus such continuous processes cannot explain the elevated levels of desorbed ices at the hotspot.

One possibility is that desorption is caused by collisions between small gas clumps that are interacting and merging \citep{Takakuwa03,Buckle06}. In this scenario, collisions between individual grains cause a transient heating which can result in runaway recombination of free radicals in the ice and subsequent sublimation of ice mantle molecules \citep{SchutteGreenberg91}. For collision velocities of the same order as the methanol line width of $\sim$0.7 km\,s$^{-1}$, colliding grain temperatures would rise to $\gtrsim$90 K \citep[following][]{Draine85}, warm enough to even desorb water and methanol thermally. Evaporation of a small fraction of the outer ice layers in the form of H$_2$O or CH$_3$OH would rapidly cool these grains down to $<$25~K and thus be consistent with the observed low dust temperatures and gas-phase abundances of water and methanol as compared to the ice toward IRS~1.

As mentioned, the second methanol peak nearly coincides with a column density peak and temperature minimum (12~K) in continuum, indicating the presence of a core, and it is only about 0.07~pc away from IRS~1. It also exhibits a lower methanol and water abundance than the methanol hotspot -- water abundance is close to that observed toward L1544 \citep{Caselli12}. Therefore UV photodesorption of water in a layer around the core; similar to the L1544 model; cannot be excluded, but the position is also close enough to IRS~1 for propagating MHD waves to cause mantle disruption.

Can the fact that water is not detected in the other three sample sources be accounted for in this scenario? In the two TMC 1 sources the estimated gas-phase CH$_3$OH/H$_2$O ratios, based on methanol data from \citet{Takakuwa00,Takakuwa03}, are at least five times higher than in B5. Thus, the desorbed outer monolayers would have to reflect that ratio. However, if  water line self-absorption is as severe as in B5, the upper limits given in Table~\ref{Nondettab} could be adjusted upward by as much as a factor of five, accounting for that difference. The upper limit in L1512 is not significant assuming that similar mantle removal processes are at work.

The discovery of gas-phase water in B5 brings new insight to the enigma of interstellar water chemistry. In B5, future observations of other complex mantle molecules believed to form along with methanol,  as well as various deuterated isotopologues, will shed further light on the nature of the surface desorption process.

\acknowledgments

Part of this work was supported by NASA's Exobiology Program and The Goddard Center for Astrobiology. E.S.W. and C.M.P. acknowledge generous support from the Swedish National Space Board.

{\it Facility:} \facility{\textit{Herschel}}

\bibliography{references}

\newcommand{\noopsort}[1]{}
\begin{thebibliography}{41}
\expandafter\ifx\csname natexlab\endcsname\relax\def\natexlab#1{#1}\fi

\bibitem[{{Andr{\'e}} {et~al.}(2010){Andr{\'e}}, {Men'shchikov}, {Bontemps},
  {K{\"o}nyves}, {Motte}, {Schneider}, {Didelon}, {Minier}, {Saraceno},
  {Ward-Thompson}, {di Francesco}, {White}, {Molinari}, {Testi}, {Abergel},
  {Griffin}, {Henning}, {Royer}, {Mer{\'{\i}}n}, {Vavrek}, {Attard},
  {Arzoumanian}, {Wilson}, {Ade}, {Aussel}, {Baluteau}, {Benedettini},
  {Bernard}, {Blommaert}, {Cambr{\'e}sy}, {Cox}, {di Giorgio}, {Hargrave},
  {Hennemann}, {Huang}, {Kirk}, {Krause}, {Launhardt}, {Leeks}, {Le Pennec},
  {Li}, {Martin}, {Maury}, {Olofsson}, {Omont}, {Peretto}, {Pezzuto}, {Prusti},
  {Roussel}, {Russeil}, {Sauvage}, {Sibthorpe}, {Sicilia-Aguilar}, {Spinoglio},
  {Waelkens}, {Woodcraft}, \& {Zavagno}}]{Andre10}
{Andr{\'e}}, P., {Men'shchikov}, A., {Bontemps}, S., {et~al.} 2010, \aap, 518,
  L102

\bibitem[{{Ashby} {et~al.}(2000){Ashby}, {Bergin}, {Plume}, {Carpenter},
  {Neufeld}, {Chin}, {Erickson}, {Goldsmith}, {Harwit}, {Howe}, {Kleiner},
  {Koch}, {Patten}, {Schieder}, {Snell}, {Stauffer}, {Tolls}, {Wang},
  {Winnewisser}, {Zhang}, \& {Melnick}}]{Ashby00}
{Ashby}, M.~L.~N., {Bergin}, E.~A., {Plume}, R., {et~al.} 2000, \apjl, 539,
  L115

\bibitem[{{Boogert} {et~al.}(2004){Boogert}, {Pontoppidan}, {Lahuis},
  {J{\o}rgensen}, {Augereau}, {Blake}, {Brooke}, {Brown}, {Dullemond}, {Evans},
  {Geers}, {Hogerheijde}, {Kessler-Silacci}, {Knez}, {Morris},
  {Noriega-Crespo}, {Sch{\"o}ier}, {van Dishoeck}, {Allen}, {Harvey},
  {Koerner}, {Mundy}, {Myers}, {Padgett}, {Sargent}, \&
  {Stapelfeldt}}]{Boogert04}
{Boogert}, A.~C.~A., {Pontoppidan}, K.~M., {Lahuis}, F., {et~al.} 2004, \apjs,
  154, 359

\bibitem[{{Buckle} {et~al.}(2006){Buckle}, {Rodgers}, {Wirstrom}, {Charnley},
  {Markwick-Kemper}, {Butner}, \& {Takakuwa}}]{Buckle06}
{Buckle}, J.~V., {Rodgers}, S.~D., {Wirstrom}, E.~S., {et~al.} 2006, Faraday
  Discussions, 133, 63

\bibitem[{{Caselli} {et~al.}(2012){Caselli}, {Keto}, {Bergin}, {Tafalla},
  {Aikawa}, {Douglas}, {Pagani}, {Y{\'{\i}}ld{\'{\i}}z}, {van der Tak},
  {Walmsley}, {Codella}, {Nisini}, {Kristensen}, \& {van Dishoeck}}]{Caselli12}
{Caselli}, P., {Keto}, E., {Bergin}, E.~A., {et~al.} 2012, \apjl, 759, L37

\bibitem[{{Charnley} \& {Rodgers}(2009)}]{CharnleyRodgers09}
{Charnley}, S.~B., \& {Rodgers}, S.~B. 2009, in Astronomical Society of the
  Pacific Conference Series, Vol. 420, Bioastronomy 2007: Molecules, Microbes
  and Extraterrestrial Life, ed. {K.~J.~Meech, J.~V.~Keane, M.~J.~Mumma,
  J.~L.~Siefert, \& D.~J.~Werthimer }, 29--+

\bibitem[{{Daniel} {et~al.}(2011){Daniel}, {Dubernet}, \&
  {Grosjean}}]{Daniel11}
{Daniel}, F., {Dubernet}, M.-L., \& {Grosjean}, A. 2011, \aap, 536, A76

\bibitem[{{de Graauw} {et~al.}(2010){de Graauw}, {Helmich}, {Phillips},
  {Stutzki}, {Caux}, {Whyborn}, {Dieleman}, {Roelfsema}, {Aarts}, {Assendorp},
  {Bachiller}, {Baechtold}, {Barcia}, {Beintema}, {Belitsky}, {Benz}, {Bieber},
  {Boogert}, {Borys}, {Bumble}, {Ca{\"\i}s}, {Caris}, {Cerulli-Irelli},
  {Chattopadhyay}, {Cherednichenko}, {Ciechanowicz}, {Coeur-Joly}, {Comito},
  {Cros}, {de Jonge}, {de Lange}, {Delforges}, {Delorme}, {den Boggende},
  {Desbat}, {Diez-Gonz{\'a}lez}, {di Giorgio}, {Dubbeldam}, {Edwards},
  {Eggens}, {Erickson}, {Evers}, {Fich}, {Finn}, {Franke}, {Gaier}, {Gal},
  {Gao}, {Gallego}, {Gauffre}, {Gill}, {Glenz}, {Golstein}, {Goulooze},
  {Gunsing}, {G{\"u}sten}, {Hartogh}, {Hatch}, {Higgins}, {Honingh}, {Huisman},
  {Jackson}, {Jacobs}, {Jacobs}, {Jarchow}, {Javadi}, {Jellema}, {Justen},
  {Karpov}, {Kasemann}, {Kawamura}, {Keizer}, {Kester}, {Klapwijk}, {Klein},
  {Kollberg}, {Kooi}, {Kooiman}, {Kopf}, {Krause}, {Krieg}, {Kramer},
  {Kruizenga}, {Kuhn}, {Laauwen}, {Lai}, {Larsson}, {Leduc}, {Leinz}, {Lin},
  {Liseau}, {Liu}, {Loose}, {L{\'o}pez-Fernandez}, {Lord}, {Luinge}, {Marston},
  {Mart{\'{\i}}n-Pintado}, {Maestrini}, {Maiwald}, {McCoey}, {Mehdi}, {Megej},
  {Melchior}, {Meinsma}, {Merkel}, {Michalska}, {Monstein}, {Moratschke},
  {Morris}, {Muller}, {Murphy}, {Naber}, {Natale}, {Nowosielski}, {Nuzzolo},
  {Olberg}, {Olbrich}, {Orfei}, {Orleanski}, {Ossenkopf}, {Peacock}, {Pearson},
  {Peron}, {Phillip-May}, {Piazzo}, {Planesas}, {Rataj}, {Ravera}, {Risacher},
  {Salez}, {Samoska}, {Saraceno}, {Schieder}, {Schlecht}, {Schl{\"o}der},
  {Schm{\"u}lling}, {Schultz}, {Schuster}, {Siebertz}, {Smit}, {Szczerba},
  {Shipman}, {Steinmetz}, {Stern}, {Stokroos}, {Teipen}, {Teyssier}, {Tils},
  {Trappe}, {van Baaren}, {van Leeuwen}, {van de Stadt}, {Visser}, {Wildeman},
  {Wafelbakker}, {Ward}, {Wesselius}, {Wild}, {Wulff}, {Wunsch}, {Tielens},
  {Zaal}, {Zirath}, {Zmuidzinas}, \& {Zwart}}]{deGraauw10}
{de Graauw}, T., {Helmich}, F.~P., {Phillips}, T.~G., {et~al.} 2010, \aap, 518,
  L6

\bibitem[{{Draine}(1985)}]{Draine85}
{Draine}, B.~T. 1985, in Protostars and Planets II, ed. D.~C. {Black} \& M.~S.
  {Matthews}, 621--640

\bibitem[{{Dubernet} {et~al.}(2009){Dubernet}, {Daniel}, {Grosjean}, \&
  {Lin}}]{Dubernet09}
{Dubernet}, M.-L., {Daniel}, F., {Grosjean}, A., \& {Lin}, C.~Y. 2009, \aap,
  497, 911

\bibitem[{{Flower} {et~al.}(2006){Flower}, {Pineau Des Forets}, \&
  {Walmsley}}]{Flower06}
{Flower}, D.~R., {Pineau Des Forets}, G., \& {Walmsley}, C.~M. 2006, \aap, 449,
  621

\bibitem[{{Garrod} \& {Pauly}(2011)}]{GarrodPauly11}
{Garrod}, R.~T., \& {Pauly}, T. 2011, \apj, 735, 15

\bibitem[{{Goldsmith} {et~al.}(1986){Goldsmith}, {Langer}, \&
  {Wilson}}]{Goldsmith86}
{Goldsmith}, P.~F., {Langer}, W.~D., \& {Wilson}, R.~W. 1986, \apjl, 303, L11

\bibitem[{{Hatchell} {et~al.}(2005){Hatchell}, {Richer}, {Fuller},
  {Qualtrough}, {Ladd}, \& {Chandler}}]{Hatchell05b}
{Hatchell}, J., {Richer}, J.~S., {Fuller}, G.~A., {et~al.} 2005, \aap, 440, 151

\bibitem[{{Klotz} {et~al.}(2008){Klotz}, {Harju}, {Ristorcelli}, {Juvela},
  {Boudet}, \& {Haikala}}]{Klotz08}
{Klotz}, A., {Harju}, J., {Ristorcelli}, I., {et~al.} 2008, \aap, 488, 559

\bibitem[{{Kristensen} \& {van Dishoeck}(2011)}]{KristensenDishoeck11}
{Kristensen}, L.~E., \& {van Dishoeck}, E.~F. 2011, Astronomische Nachrichten,
  332, 475

\bibitem[{{Langer} {et~al.}(1996){Langer}, {Velusamy}, \& {Xie}}]{Langer96}
{Langer}, W.~D., {Velusamy}, T., \& {Xie}, T. 1996, \apjl, 468, L41

\bibitem[{{Leger} {et~al.}(1985){Leger}, {Jura}, \& {Omont}}]{Leger85}
{Leger}, A., {Jura}, M., \& {Omont}, A. 1985, \aap, 144, 147

\bibitem[{{Maercker} {et~al.}(2008){Maercker}, {Sch{\"o}ier}, {Olofsson},
  {Bergman}, \& {Ramstedt}}]{Maercker08}
{Maercker}, M., {Sch{\"o}ier}, F.~L., {Olofsson}, H., {Bergman}, P., \&
  {Ramstedt}, S. 2008, \aap, 479, 779

\bibitem[{{Markwick} {et~al.}(2000){Markwick}, {Millar}, \&
  {Charnley}}]{Markwick00}
{Markwick}, A.~J., {Millar}, T.~J., \& {Charnley}, S.~B. 2000, \apj, 535, 256

\bibitem[{{M{\"u}ller} {et~al.}(2005){M{\"u}ller}, {Schl{\"o}der}, {Stutzki},
  \& {Winnewisser}}]{Muller05}
{M{\"u}ller}, H.~S.~P., {Schl{\"o}der}, F., {Stutzki}, J., \& {Winnewisser}, G.
  2005, Journal of Molecular Structure, 742, 215

\bibitem[{{{\"O}berg} {et~al.}(2011){{\"O}berg}, {Boogert}, {Pontoppidan}, {van
  den Broek}, {van Dishoeck}, {Bottinelli}, {Blake}, \& {Evans}}]{Oberg11}
{{\"O}berg}, K.~I., {Boogert}, A.~C.~A., {Pontoppidan}, K.~M., {et~al.} 2011,
  \apj, 740, 109

\bibitem[{{Ott}(2010)}]{Ott10}
{Ott}, S. 2010, in Astronomical Society of the Pacific Conference Series, Vol.
  434, Astronomical Data Analysis Software and Systems XIX, ed. Y.~{Mizumoto},
  K.-I. {Morita}, \& M.~{Ohishi}, 139

\bibitem[{{Pagani} {et~al.}(2011){Pagani}, {Roueff}, \& {Lesaffre}}]{Pagani11}
{Pagani}, L., {Roueff}, E., \& {Lesaffre}, P. 2011, \apjl, 739, L35

\bibitem[{{Pilbratt} {et~al.}(2010){Pilbratt}, {Riedinger}, {Passvogel},
  {Crone}, {Doyle}, {Gageur}, {Heras}, {Jewell}, {Metcalfe}, {Ott}, \&
  {Schmidt}}]{Pilbratt10}
{Pilbratt}, G.~L., {Riedinger}, J.~R., {Passvogel}, T., {et~al.} 2010, \aap,
  518, L1

\bibitem[{{Pineda} {et~al.}(2011){Pineda}, {Goodman}, {Arce}, {Caselli},
  {Longmore}, \& {Corder}}]{Pineda11}
{Pineda}, J.~E., {Goodman}, A.~A., {Arce}, H.~G., {et~al.} 2011, \apjl, 739, L2

\bibitem[{{Prasad} \& {Tarafdar}(1983)}]{PrasadTarafdar83}
{Prasad}, S.~S., \& {Tarafdar}, S.~P. 1983, \apj, 267, 603

\bibitem[{{Rabli} \& {Flower}(2010)}]{RabliFlower10}
{Rabli}, D., \& {Flower}, D.~R. 2010, \mnras, 406, 95

\bibitem[{{Roelfsema} {et~al.}(2012){Roelfsema}, {Helmich}, {Teyssier},
  {Ossenkopf}, {Morris}, {Olberg}, {Shipman}, {Risacher}, {Akyilmaz},
  {Assendorp}, {Avruch}, {Beintema}, {Biver}, {Boogert}, {Borys}, {Braine},
  {Caris}, {Caux}, {Cernicharo}, {Coeur-Joly}, {Comito}, {de Lange},
  {Delforge}, {Dieleman}, {Dubbeldam}, {de Graauw}, {Edwards}, {Fich},
  {Flederus}, {Gal}, {di Giorgio}, {Herpin}, {Higgins}, {Hoac}, {Huisman},
  {Jarchow}, {Jellema}, {de Jonge}, {Kester}, {Klein}, {Kooi}, {Kramer},
  {Laauwen}, {Larsson}, {Leinz}, {Lord}, {Lorenzani}, {Luinge}, {Marston},
  {Mart{\'{\i}}n-Pintado}, {McCoey}, {Melchior}, {Michalska}, {Moreno},
  {M{\"u}ller}, {Nowosielski}, {Okada}, {Orlea{\'n}ski}, {Phillips}, {Pearson},
  {Rabois}, {Ravera}, {Rector}, {Rengel}, {Sagawa}, {Salomons},
  {S{\'a}nchez-Su{\'a}rez}, {Schieder}, {Schl{\"o}der}, {Schm{\"u}lling},
  {Soldati}, {Stutzki}, {Thomas}, {Tielens}, {Vastel}, {Wildeman}, {Xie},
  {Xilouris}, {Wafelbakker}, {Whyborn}, {Zaal}, {Bell}, {Bjerkeli}, {De Beck},
  {Cavali{\'e}}, {Crockett}, {Hily-Blant}, {Kama}, {Kaminski}, {Lefl{\'o}ch},
  {Lombaert}, {de Luca}, {Makai}, {Marseille}, {Nagy}, {Pacheco}, {van der
  Wiel}, {Wang}, \& {Y{\i}ld{\i}z}}]{Roelfsema12}
{Roelfsema}, P.~R., {Helmich}, F.~P., {Teyssier}, D., {et~al.} 2012, \aap, 537,
  A17

\bibitem[{{Rybicki} \& {Hummer}(1991)}]{RybickiHummer91}
{Rybicki}, G.~B., \& {Hummer}, D.~G. 1991, \aap, 245, 171

\bibitem[{{Sandford} \& {Allamandola}(1993)}]{SandfordAllamandola93}
{Sandford}, S.~A., \& {Allamandola}, L.~J. 1993, \apj, 417, 815

\bibitem[{{Schnee} {et~al.}(2008){Schnee}, {Li}, {Goodman}, \&
  {Sargent}}]{Schnee08}
{Schnee}, S., {Li}, J., {Goodman}, A.~A., \& {Sargent}, A.~I. 2008, \apj, 684,
  1228

\bibitem[{{Sch{\"o}ier} {et~al.}(2005){Sch{\"o}ier}, {van der Tak}, {van
  Dishoeck}, \& {Black}}]{Schoier05}
{Sch{\"o}ier}, F.~L., {van der Tak}, F.~F.~S., {van Dishoeck}, E.~F., \&
  {Black}, J.~H. 2005, \aap, 432, 369

\bibitem[{{Schutte} \& {Greenberg}(1991)}]{SchutteGreenberg91}
{Schutte}, W.~A., \& {Greenberg}, J.~M. 1991, \aap, 244, 190

\bibitem[{{Snell} {et~al.}(2000){Snell}, {Howe}, {Ashby}, {Bergin}, {Chin},
  {Erickson}, {Goldsmith}, {Harwit}, {Kleiner}, {Koch}, {Neufeld}, {Patten},
  {Plume}, {Schieder}, {Stauffer}, {Tolls}, {Wang}, {Winnewisser}, {Zhang}, \&
  {Melnick}}]{Snell00}
{Snell}, R.~L., {Howe}, J.~E., {Ashby}, M.~L.~N., {et~al.} 2000, \apjl, 539,
  L93

\bibitem[{{Takakuwa} {et~al.}(2003){Takakuwa}, {Kamazaki}, {Saito}, \&
  {Hirano}}]{Takakuwa03}
{Takakuwa}, S., {Kamazaki}, T., {Saito}, M., \& {Hirano}, N. 2003, \apj, 584,
  818

\bibitem[{{Takakuwa} {et~al.}(2000){Takakuwa}, {Mikami}, {Saito}, \&
  {Hirano}}]{Takakuwa00}
{Takakuwa}, S., {Mikami}, H., {Saito}, M., \& {Hirano}, N. 2000, \apj, 542, 367

\bibitem[{{Troscompt} {et~al.}(2009){Troscompt}, {Faure}, {Maret},
  {Ceccarelli}, {Hily-Blant}, \& {Wiesenfeld}}]{Troscompt09}
{Troscompt}, N., {Faure}, A., {Maret}, S., {et~al.} 2009, \aap, 506, 1243

\bibitem[{{van Dishoeck} {et~al.}(2011){van Dishoeck}, {Kristensen}, {Benz},
  {Bergin}, {Caselli}, {Cernicharo}, {Herpin}, {Hogerheijde}, {Johnstone},
  {Liseau}, {Nisini}, {Shipman}, {Tafalla}, {van der Tak}, {Wyrowski},
  {Aikawa}, {Bachiller}, {Baudry}, {Benedettini}, {Bjerkeli}, {Blake},
  {Bontemps}, {Braine}, {Brinch}, {Bruderer}, {Chavarr{\'{\i}}a}, {Codella},
  {Daniel}, {de Graauw}, {Deul}, {di Giorgio}, {Dominik}, {Doty}, {Dubernet},
  {Encrenaz}, {Feuchtgruber}, {Fich}, {Frieswijk}, {Fuente}, {Giannini},
  {Goicoechea}, {Helmich}, {Herczeg}, {Jacq}, {J{\o}rgensen}, {Karska},
  {Kaufman}, {Keto}, {Larsson}, {Lefloch}, {Lis}, {Marseille}, {McCoey},
  {Melnick}, {Neufeld}, {Olberg}, {Pagani}, {Pani{\'c}}, {Parise}, {Pearson},
  {Plume}, {Risacher}, {Salter}, {Santiago-Garc{\'{\i}}a}, {Saraceno},
  {St{\"a}uber}, {van Kempen}, {Visser}, {Viti}, {Walmsley}, {Wampfler}, \&
  {Y{\i}ld{\i}z}}]{Dishoecketal11}
{van Dishoeck}, E.~F., {Kristensen}, L.~E., {Benz}, A.~O., {et~al.} 2011,
  \pasp, 123, 138

\bibitem[{{Vasyunin} \& {Herbst}(2013)}]{VasyuninHerbst13a}
{Vasyunin}, A.~I., \& {Herbst}, E. 2013, \apj, 762, 86

\bibitem[{{Yu} {et~al.}(1999){Yu}, {Billawala}, \& {Bally}}]{Yu99}
{Yu}, K.~C., {Billawala}, Y., \& {Bally}, J. 1999, \aj, 118, 2940

\end{thebibliography}

\clearpage

\clearpage

\begin{table} 
\begin{center}
\caption{Summary of Non-detection Observations \label{Nondettab}}
\begin{tabular}{l l l l l l}
\tableline\tableline
Source &ObsId &$T_{\rm rms}$\tablenotemark{a}  & $X$($o$-H$_2$O) \\
\tableline
\rule{0pt}{2ex}L1512\_meth &1342250434	&37 mK&$<$1.5e-8\\
TMC-1C, clump6-1 		&1342266589&	9 mK&	$<$4e-9\\
TMC-1, ch3oh-1			&1342266590&	8 mK&	$<$4e-9\\
\tableline
\end{tabular}
\tablenotetext{a}{At 0.3~km\,s$^{-1}$ resolution.}
\end{center}
\end{table}

\begin{table} 
\begin{center}
\caption{ALI Cloud Model Properties \label{ALItab}}
\begin{tabular}{l c c c}
\tableline\tableline
& CH$_3$OH & $o-$H$_2$O & $o-$H$_2$O\\ 
 & "$hotspot$" & "$hotspot$" & $2nd\,B5\,pos$ \\
\tableline
\rule{0pt}{2ex} $Clump$\\
$R_{\rm max}$ [pc]			&0.04	&0.04 	&0.04\\
$n$(H$_2$) [cm$^{-3}$]		&5.0(4)	&5.0(4)	&9.0(4)\\
$M$ [$M_{\sun}$]			&0.64	&0.64	&1.35\\
$T_{\rm kin}$ [K] 			&9		&9		&10\\
$T_{\rm dust}$ [K] 			&13		&13		&14\\
$v_{\rm turb}$ [km\,s$^{-1}$]	&0.3 		&0.42	&0.5\\
$v_{\rm exp,max}$[km\,s$^{-1}$]	&-		&0.12	&0.5\\
$X_{\rm mol}$				&3.9($-8$)	&1.9($-8$)	&5.5($-9$)\\
\rule{0pt}{2ex} $Envelope$ \\
$R_{\rm max}$ [pc]			&-		&0.08	&0.08\\
$n$(H$_2$) [cm$^{-3}$]		&- 		&5.0(3)	&9.0(3)\\
$M$ [$M_{\sun}$]			&- 		&0.55	&0.79\\
$v_{\rm exp}$ [km\,s$^{-1}$]	&- 		&0.12	&0.5\\
\tableline
\end{tabular}
\tablecomments{5.0(4) means 5.0$\times$10$^4$, etc.}
\end{center}
\end{table}

\end{document}